# Quantized Precoding for Multi-Antenna Downlink Channels with MAGIQ

Andrei Nedelcu*, Fabian Steiner*, Markus Staudacher*, Gerhard Kramer*,
Wolfgang Zirwas†, Rakash Sivasiva Ganesan†, Paolo Baracca‡, Stefan Wesemann‡
*Institute for Communications Engineering, Technical University of Munich, Germany
†Nokia Bell Labs, Munich, Germany,   ‡Nokia Bell Labs, Stuttgart, Germany

*Abstract*—A multi-antenna, greedy, iterative, and quantized (MAGIQ) precoding algorithm is proposed for downlink channels. MAGIQ allows a straightforward integration with orthogonal frequency-division multiplexing (OFDM) for frequency selective channels. MAGIQ is compared to three existing algorithms in terms of information rates and complexity: quantized linear precoding (QLP), SQUID, and an ADMM-based algorithm. The information rate is measured by using a lower bound for finite modulation sets, and the complexity is measured by the number of multiplications and comparisons. MAGIQ and ADMM achieve similar information rates for Rayleigh flat-fading channels and one-bit quantization per real dimension, and they outperform QLP and SQUID for higher order modulation formats.

## I. INTRODUCTION

Massive multiple-input multiple-output (massive MIMO) uses large antenna arrays at base stations to serve many users that each have a small number of antennas [1]. The gains of massive MIMO include improved power and spectral efficiencies, and simplified signal processing [2]. The gains are often stated for a large number $N$ of base station antennas and a large number $K$ of user equipments (UEs) when the ratio $N/K$ is held constant.

A practical implementation for large $N$ and $K$ is challenging. For example, consider a base station deployment with $N$ radio frequency (RF) chains. It seems impractical to use high-resolution analog-to-digital and digital-to-analog converters (ADCs/DACs), along with linear but low-efficiency power amplifiers. Two potential solutions are as follows. First, hybrid-beamforming [3] uses analog beamformers in the RF chain of each antenna, and the digital baseband processing is shared among different RF chains. Second, low-resolution ADCs/DACs simplify the transceivers, e.g., one bit quantizers use simple comparators and they might permit using non-linear power amplifiers.

### A. Uplink and Downlink

We consider the low-resolution quantizer approach. There are numerous studies on the *uplink* with either linear detectors, e.g., matched filter (MF), zero forcing (ZF), and Wiener filter (WF), or non-linear detectors such as approximate message passing (AMP) [4]–[6]. For example, it is known that even for low-resolution quantization at the base station antennas, the UEs can communicate with higher order modulation formats if $N$ is sufficiently large.

Recently, the *downlink* has received attention [7]–[12]. The authors of [7] use the Bussgang approximation to design quantized linear precoders (QLPs). The paper [8] introduces a lookup-table based precoder for quadrature phase-shift keying (QPSK) that minimizes the uncoded bit error rate (BER) at the UE. The paper [9] introduces a hybrid RF architecture that combines a constrained and a conventional MIMO array. A greedy knapsack-like algorithm is used to minimize the mean square error (MSE) between the desired and constructed UE signal points.

The authors of [10] describe nonlinear approaches based on semidefinite relaxation, $\ell_\infty$ norm relaxation and sphere decoding. For a reasonable performance and complexity trade-off, they recommend $\ell_\infty$ norm relaxation, which is named SQUID [10, Sec. IV]. Another approach is described in [11], where the authors extend the framework of alternating direction of multipliers (ADMM), and they report slight improvements over SQUID. The precoding problem for coarsely quantized, frequency-selective channels and its integration with OFDM was considered in [12], where the authors use linear precoding and a frequency domain approach. An extension of SQUID to OFDM and frequency selective channels was presented in [13].

### B. Contributions and Organization

Our work is inspired by the greedy approach of [9]. We introduce a multi-antenna, greedy, iterative, and quantized (MAGIQ) precoding algorithm for downlink channels. The algorithm decomposes a high dimensional nonlinear and non-convex problem into low dimensional sub-problems that can be solved efficiently. We compare MAGIQ's performance to QLP, SQUID, and the ADMM-based algorithm [7], [10]–[12]. We consider both low-order and higher-order modulation formats (4, 16, 64-quadrature amplitude modulation (QAM)). As a key performance metric, we compute lower bounds on the information rates by using the mismatched decoding framework [14], [15]. We thus take modulation constraints into account, rather than idealized Gaussian signaling. For the frequency-selective case, our approach operates in the time domain and avoids switching between domains to enforce the discrete alphabet constraint. A related problem was presented in the context of precoding for frequency selective channels with constant envelope continuous modulation in [16].

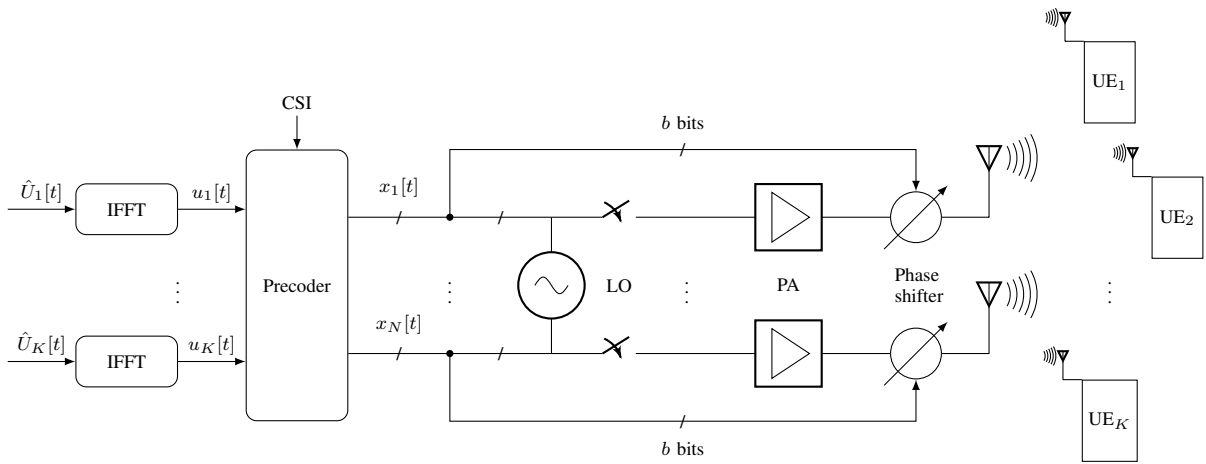

Fig. 1. Multi-user MIMO downlink with a low resolution digitally controlled analog architecture.

This paper is organized as follows. In Sec. II, we describe the system model, the precoding problem for a $K$-user downlink channel with coarsely quantized transmit symbols and the QLP approaches. In Sec. III, we present the MAGIQ precoding algorithm. Sec. IV provides a comprehensive complexity comparison of SQUID, ADMM and MAGIQ in terms of arithmetic operations. Sec. V presents simulation results and we conclude in Sec. VI.

## II. Preliminaries

### A. System Model

Consider the downlink of a multi-user MIMO channel with $N$ transmit antennas and $K$ UEs that each have a single antenna. A discrete time, frequency selective, baseband channel has a finite impulse response (FIR) filter between each pair of transmit and receive antennas. We collect the received signals $y_k[t]$ of user $k$, $k = 1, 2, \ldots, K$, at time $t$ into the $K$-dimensional column vector

$$\boldsymbol{y}[t] = \sum_{l=0}^{L-1} \boldsymbol{H}[l]\boldsymbol{x}[t-l] + \boldsymbol{z}[t] \quad (1)$$

where the noise $\boldsymbol{z}[t]$ is a circularly-symmetric, complex, Gaussian, random, column vector with a scaled identity covariance matrix, i.e., we have $\boldsymbol{z} \sim \mathcal{CN}(\boldsymbol{0}, \sigma^2 \boldsymbol{I})$. The transmit column vector $\boldsymbol{x}[t] = [x_1[t], x_2[t], \ldots, x_N[t]]^\mathrm{T}$ has entries taken from a discrete alphabet $\mathcal{X}$ that has $2^b+1$ values where $b$ bits encode the phase. More precisely, we choose

$$\mathcal{X} = \{0, \exp(\mathrm{j}\, 2\pi q/2^b)\} \quad (2)$$

with $q = 0, 1, \ldots, 2^b - 1$. We choose the transmit power to be at most unity, and we define the transmit signal-to-noise ratio (SNR) as $\mathrm{SNR} = 1/\sigma^2$. The channel impulse response matrix is

$$\boldsymbol{H}[l] = \begin{pmatrix} h_{11}[l] & h_{12}[l] & \ldots & h_{1N}[l] \\ h_{21}[l] & h_{22}[l] & \ldots & h_{2N}[l] \\ \vdots & \vdots & \ddots & \vdots \\ h_{K1}[l] & h_{K2}[l] & \ldots & h_{KN}[l] \end{pmatrix} \quad (3)$$

where $h_{kn}[l]$, $l = 0, 1, \ldots, L-1$, is the channel impulse response from the $n$-th antenna at the base station to the $k$-th UE. We study a Rayleigh fading frequency selective channel with a uniform power delay profile, i.e., we choose $\mathrm{E}\left[|h_{kn}[l]|^2\right] = 1/L$, where the $h_{kn}[l] \sim \mathcal{CN}(0, 1/L)$ are i.i.d. circularly-symmetric, complex Gaussian random variables. We further assume a block fading channel model, i.e., the channel realization remains constant for the coherence interval, and the instantaneous realizations $\boldsymbol{H}[l]$, $l = 0, 1, \ldots, L-1$, are known at the transmitter.

We remark that the alphabet (2) can be interpreted as permitting per-symbol antenna selection through the zero symbol. The idea of joint precoding and antenna selection also appeared in [17], but our algorithm selects antennas without an extra metric that enforces sparsity.

### B. Flat Fading Channels

For flat fading channels ($L = 1$) the channel (1) becomes

$$\boldsymbol{y} = \boldsymbol{H}\boldsymbol{x} + \boldsymbol{z}. \quad (4)$$

Let $u_k \in \mathcal{U}$ be the noise-free complex symbol that we wish to generate at the $k$-th UE for $k = 1, \ldots, K$, where $\mathcal{U}$ is either a 4-, 16-, or 64-QAM signaling set. Let $\boldsymbol{u}$ be the column vector with the $K$ symbols. Consider the precoding problem

$$\begin{aligned} \min_{\boldsymbol{x},\alpha} \quad & \|\boldsymbol{u} - \alpha \boldsymbol{H}\boldsymbol{x}\|_2^2 + \alpha^2 K \sigma^2 \\ \text{s.t.} \quad & \boldsymbol{x} \in \mathcal{X}^N \\ & \alpha > 0. \end{aligned} \quad (5)$$

The factor $\alpha$ permits trading off noise enhancement and the received signal power, the latter being more important at low SNR and maximized by the MF [?]. For a fixed value of $\alpha$, the problem (5) represents a classic nonlinear integer least-squares problem [18].

### C. Quantized Linear Precoding

QLP approximates the solution of (5) by $\boldsymbol{x} = \mathrm{Q}(\boldsymbol{P}\boldsymbol{u})$, where $\boldsymbol{P} \in \mathbb{C}^{N \times K}$ is a precoding matrix and $\mathrm{Q}(\cdot)$ is a

**Algorithm 1:** MAGIQ for frequency-flat channels

1 **Inputs**: $\boldsymbol{u}, \boldsymbol{H}, \mathcal{S} = \{1, \ldots, N\}, err_{\min}$
2 **Initialize**: $\boldsymbol{x} = \boldsymbol{x}_{init}, \alpha = \alpha_{init}$
3 **for** $iter = 1 : I$ **do**
4 $\quad loopIdx = 1$
5 $\quad err_0 = \|\boldsymbol{u}\|^2$
6 $\quad$ **while** $(err_{loopIdx} > err_{\min}) \vee (err_{loopIdx} < err_{loopIdx-1}) \vee (loopIdx \leq N)$ **do**
7 $\quad\quad (x_{n^\star}^\star, n^\star) = \operatorname{argmin}_{x_n \in \mathcal{X}, n \in \mathcal{S}} F(\boldsymbol{x}, \alpha)$
8 $\quad\quad (x_1, \ldots, x_n, \ldots, x_N)^{\mathrm{T}} = (x_1, \ldots, x_{n^\star}^\star, \ldots, x_N)^{\mathrm{T}}$
9 $\quad\quad \mathcal{S} \leftarrow \mathcal{S} \setminus \{n^\star\}$
10 $\quad\quad loopIdx \leftarrow loopIdx + 1$
11 $\quad\quad err_{loopIdx} = \|\boldsymbol{u} - \alpha \boldsymbol{H}\boldsymbol{x}\|_2^2 + \alpha^2 K \sigma^2$
12 $\quad$ **end**
13 $\quad \alpha = \frac{\operatorname{Re}(\boldsymbol{u}^{\mathrm{H}} \boldsymbol{H} \boldsymbol{x})}{\|\boldsymbol{H}\boldsymbol{x}\|_2^2 + K\sigma^2}$
14 **end**
15 **Output** $\boldsymbol{x}, \alpha$

quantization function with range $\mathcal{X}^N$ that operates entry-wise. Common choices for $\boldsymbol{P}$ are MF and ZF precoders, namely the respective

$$\boldsymbol{P} = \frac{1}{\alpha_{\mathrm{MF}}} \boldsymbol{H}^{\mathrm{H}} \qquad \alpha_{\mathrm{MF}} = \operatorname{tr}\left(\boldsymbol{H}\boldsymbol{H}^{\mathrm{H}}\right), \quad (6)$$

$$\boldsymbol{P} = \frac{1}{\alpha_{\mathrm{ZF}}} \boldsymbol{H}^{\mathrm{H}}(\boldsymbol{H}\boldsymbol{H}^{\mathrm{H}})^{-1} \quad \alpha_{\mathrm{ZF}} = \operatorname{tr}\left((\boldsymbol{H}\boldsymbol{H}^{\mathrm{H}})^{-1}\right). \quad (7)$$

QLP is conceptually simple, but it performs poorly for higher order modulations and intermediate ranges of $N$, as shown below.

### III. MAGIQ PRECODING

#### A. Frequency-Flat Channels

Algorithm 1 outlines MAGIQ for frequency-flat channels ($L = 1$). The inner loop evaluates the cost function

$$F(\boldsymbol{x}, \alpha) = \|\boldsymbol{u} - \alpha \boldsymbol{H}\boldsymbol{x}\|_2^2 + \alpha^2 K \sigma^2 \quad (8)$$

and selects (without replacement) an antenna index $n$ and a precoded symbol $x_n$ from $\mathcal{X}$ such that (8) is minimized. That is, each iteration selects the best coordinate in a greedy fashion. After forming a precoded vector $\boldsymbol{x}$, the factor $\alpha$ is chosen to make the gradient of (5) with respect to $\alpha$ become zero. The factor $\alpha$ is thus a function of the channel, the target symbol $\boldsymbol{u}$, as well as the precoding vector $\boldsymbol{x}$.

The algorithm then performs iterations by updating $\boldsymbol{x}$ and $\alpha$ until it reaches a stopping criterion or a maximum number of iterations. Simulations presented in Sec. V show that the algorithm converges to a good local minimum with a small number of iterations, and that the quality of the local minimum is at least as good as the minima obtained by the SQUID and ADMM algorithms in [10], [11].

#### B. Frequency Selective Channels

For frequency selective channels ($L > 1$), the precoder puts out a string of column vectors $\boldsymbol{x}[1], \ldots, \boldsymbol{x}[T]$, each with alphabet $\mathcal{X}^N$, where $T$ is at most the coherence time of the channel. In practice, $T$ is chosen to balance the need for accurate channel state information (CSI) and quality-of-service (QoS) requirements.

Consider the $K$-dimensional column vectors $\boldsymbol{u}[1], \ldots, \boldsymbol{u}[T]$ that we would like to generate at the $K$ UEs. We state our optimization problem as follows.

$$\min_{\boldsymbol{x}[1], \ldots, \boldsymbol{x}[T], \alpha} \sum_{t=1}^{T} \left\| \boldsymbol{u}[t] - \alpha \sum_{l=0}^{L-1} \boldsymbol{H}[l]\boldsymbol{x}[t-l] \right\|_2^2 + \alpha^2 T K \sigma^2$$

$$\text{s.t.} \quad \boldsymbol{x}[t] \in \mathcal{X}^N, \ t = 1, \ldots, T$$

$$\alpha > 0. \quad (9)$$

The optimization problem (9) suggests a time domain approach rather than the frequency domain approach of [13]. The main advantage of the former approach is that one does not need to switch between time and frequency domains to enforce the discrete alphabet constraint (2). The cost of each such switch is a length $T$ discrete Fourier transform (DFT).

#### C. Precoding for OFDM

Fig. 1 shows how OFDM can be combined with MAGIQ. The frequency domain vectors $\hat{\boldsymbol{U}}[\cdot]$ corresponding to the $K$ users are converted to the time domain vectors $\boldsymbol{u}[\cdot]$ by a length $T$ IDFT:

$$u_k[t] = \frac{1}{T} \sum_{v=1}^{T} \hat{U}_k[v] e^{\mathrm{j}\, 2\pi(v-1)(t-1)/T} \quad (10)$$

$$\hat{\boldsymbol{U}}[t] = [\hat{U}_1[t], \ldots, \hat{U}_K[t]]^{\mathrm{T}}$$

$$\boldsymbol{u}[t] = [u_1[t], \ldots, u_K[t]]^{\mathrm{T}}$$

for $k = 1, \ldots, K$, $t = 1, \ldots, T$. For the simulations, we generated the frequency domain symbols $\hat{U}_k[t]$ uniformly from QPSK, 16-QAM, and 64-QAM constellations. Each UE performs single-user OFDM demodulation followed by a hard or soft decision detection. MAGIQ is flexible with respect to shaping constraints, constellation size, number of users, number of sub-carriers, and channel models.

#### D. Algorithm Description

For frequency selective channels, the vector $\boldsymbol{x}[t]$ of transmit symbols at time $t$ should be chosen as a function of the transmit symbols at other time instances due to the channel memory, i.e., $\boldsymbol{x}[t]$ influences the channel output at times $t+1, t+2, \ldots, t+L-1$. However, a joint optimization over strings of length $T$ seems difficult because of the exponential increase in the size of the constraint space.

We approach the problem by splitting the joint optimization into a set of sub-problems with reduced complexity, see Algorithm 2. This approach is related to coordinate descent algorithms that have been successful in, e.g., compressed sensing [19]. We perform a two-fold coordinate-wise splitting

**Algorithm 2:** MAGIQ for frequency selective channels

1 **Input:** $\boldsymbol{H}[l]$, $\boldsymbol{u}[t]$, $t = 1, \ldots, T$
2 **Initialization:** $\boldsymbol{x}[t]^0 = \boldsymbol{x}[t]_{init}$, $t = 1, \ldots, T$, $\alpha = \alpha_{init}$
3 **for** $iter = 1 : \text{I}$ **do**
4    **for** $t = 1 : T$ **do**
5       $\boldsymbol{x}[t]^{iter} = \arg\min_{\tilde{\boldsymbol{x}}} G(\boldsymbol{x}^{iter}[1], \ldots, \boldsymbol{x}^{iter}[t-1], \tilde{\boldsymbol{x}},$
        $\boldsymbol{x}^{iter-1}[t+1], \ldots, \boldsymbol{x}^{iter-1}[T])$
6    **end**
7    $\alpha = \frac{\sum_{t=1}^{T} \text{Re}(\boldsymbol{u}[t]^{\text{H}} \sum_{l=0}^{L-1} \boldsymbol{H}[l]\boldsymbol{x}[t-l])}{\sum_{t=1}^{T} \left\| \sum_{l=0}^{L-1} \boldsymbol{H}[l]\boldsymbol{x}[t-l] \right\|_2^2 + TK\sigma^2}$
8 **end**
9 **Output:** $\boldsymbol{x}[t]$, $t = 1, \ldots, T, \alpha$

of the problem stated in (9). First, we solve the precoding problem for one time coordinate $t$ at a time, starting at time 1 and ending at time $T$. Under this formulation, we replace the cost function (8) with:

$$G(\boldsymbol{x}[1], \ldots, \boldsymbol{x}[t-1], \boldsymbol{x}[t], \boldsymbol{x}[t+1], \ldots, \boldsymbol{x}[T])$$
$$= \sum_{t=1}^{T} \left\| \boldsymbol{u}[t] - \alpha \sum_{l=0}^{L-1} \boldsymbol{H}[l]\boldsymbol{x}[t-l] \right\|_2^2 + \alpha^2 TK\sigma^2$$
$$= \sum_{t=1}^{T} \| \widetilde{\boldsymbol{u}}[t] - \alpha \boldsymbol{H}[0]\boldsymbol{x}[t] \|_2^2 + \alpha^2 TK\sigma^2 \quad (11)$$

where

$$\widetilde{\boldsymbol{u}}[t] = \boldsymbol{u}[t] - \alpha \sum_{l=1}^{L-1} \boldsymbol{H}[l]\boldsymbol{x}[t-l]. \quad (12)$$

The last line in (11) has the same form as the cost function in (5). We again split this problem in a coordinate-wise fashion and solve it with Algorithm 1. Algorithm 2 then iterates over the frame until prescribed convergence criteria are met. Simulations show that the performance over frequency selective channels is close to that of the best known precoders over frequency flat channels with as little as 4 iterations.

*E. Block Flat Fading*

MAGIQ precoding for flat fading in Sec. III-A has the transmitter compute $\boldsymbol{x}$ and $\alpha$ for each transmit symbol $\boldsymbol{u}$. For block flat fading, the transmitter might be able to broadcast the precoding factors to the receivers, as they do not change inside the block. However, such a broadcast channel is not necessarily available, and the number of distinct $\alpha$ values is generally large for a large number $K$ of users and for large modulation sets. In [20], the authors discuss block fading and different estimation strategies at the receiver.

We approach the problem by studying various scenarios.

- Scenario 1: The base station computes a pair $(\alpha, \boldsymbol{x})$ for each symbol $\boldsymbol{u}$, and broadcasts all precoding factors to the receivers. This scenario seems unrealistic but provides a benchmark for performance.
- Scenario 2: The base station computes a pair $(\alpha, \boldsymbol{x})$ for each symbol $\boldsymbol{u}$, and the receiver estimates an auxiliary channel as described in Sec. III-F below. This scenario requires no side information on the precoding factors.
- Scenario 3: The base station computes one $\alpha$ per coherence interval, and the receiver estimates an auxiliary channel as described in Sec. III-F below. To compute $\alpha$, the base station tries to solve the optimization problem:

$$\min_{\boldsymbol{x}[1], \ldots, \boldsymbol{x}[T], \alpha} \sum_{t=1}^{T} \|\boldsymbol{u}[t] - \alpha \boldsymbol{H}\boldsymbol{x}[t]\|_2^2 + \alpha^2 TK\sigma^2$$
$$\text{s.t.} \quad \boldsymbol{x}[t] \in \mathcal{X}^N, \ t = 1, \ldots, T \quad (13)$$
$$\alpha > 0.$$

Fig. 9 below shows that the performance of MAGIQ for Scenario 3 is close to that of Scenario 1.

*F. Information Rates*

We use the mismatched decoding framework [14], [15] to calculate lower bounds on information rates. Consider a channel $p_{Y|X}(\cdot|\cdot)$ with input $X$ and output $Y$. A lower bound to the mutual information

$$\text{I}(X;Y) = \text{E}\left[\log_2\left(\frac{p_{Y|X}(Y|X)}{\sum_a p_{Y|X}(Y|a)P_X(a)}\right)\right] \quad (14)$$

is the generalized mutual information (GMI)

$$\max_{s \geq 0} \quad \text{E}\left[\log_2\left(\frac{q_{Y|X}(Y|X)^s}{\sum_a q_{Y|X}(Y|a)^s P_X(a)}\right)\right] \quad (15)$$

where $q_{Y|X}(\cdot|\cdot)$ is an *auxiliary* channel. The lower bound is tight if $p_{Y|X} = q_{Y|X}$, but often $p_{Y|X}$ is difficult to characterize, and hence we resort to simpler channels $q_{Y|X}$.

To benchmark the performance of different precoding strategies, we consider Scenario 1 where the receivers know the base station's $\alpha$ values. As mentioned above, this scenario may be unrealistic, but we will see that the performance trends for Scenarios 2 and 3 are similar. We proceed as follows:

1) Perform Monte Carlo simulations of (1) to collect sample sequences $\boldsymbol{u}^{(1)}, \ldots, \boldsymbol{u}^{(S)}$ and $\boldsymbol{y}^{(1)}, \ldots, \boldsymbol{y}^{(S)}$ of length $S$.
2) We distinguish the different cases.
    - For Scenario 1, the receiver knows the base station's $\alpha$ values, and it scales each received sample with the corresponding $\alpha$, i.e., we compute

$$\tilde{\boldsymbol{y}}^{(i)} = \alpha^{(i)} \boldsymbol{y}^{(i)}, \quad i = 1, \ldots, S. \quad (16)$$

    - For Scenarios 2 and 3, the receivers do not know the $\alpha$ values. Instead, the effect of the $\alpha$ values is captured in the calculation of the channel, see step 3 below, and we set $\tilde{\boldsymbol{y}}^{(i)} = \boldsymbol{y}^{(i)}$.
    - For the frequency selective case of Algorithm 2, the precoding factor is updated once per coherence interval. As before, we have $\tilde{\boldsymbol{y}}[t]^{(i)} = \boldsymbol{y}[t]^{(i)}, t = 1, \ldots, T$, and the calculated channel in step 3 below is used for all received samples within one coherence interval.

3) Every receiver chooses a Gaussian auxiliary channel $\tilde{Y} = h \cdot U + Z$ with conditional density

$$q_{\tilde{Y}|U}(\tilde{y}|u; h, \sigma_q^2) = \frac{1}{\pi \sigma_q^2} e^{-\frac{|\tilde{y} - h \cdot u|^2}{\sigma_q^2}}. \quad (17)$$

The parameters $h \in \mathbb{C}$ and $\sigma_q^2 \in \mathbb{R}^+$ are obtained by maximum-likelihood (ML) estimation from the sample sequences for a particular user $k$:

$$h = \frac{\sum_{i=1}^{S} \tilde{y}_k^{(i)} u_k^{(i)*}}{\sum_{i=1}^{S} \left|u_k^{(i)}\right|^2}; \quad \sigma_q^2 = \frac{1}{S} \sum_{i=1}^{S} \left|\tilde{y}_k^{(i)} - h u_k^{(i)}\right|^2. \quad (18)$$

A Gaussian auxiliary channel seems reasonable, and it is the right choice when $N \to \infty$ by the weak law of large numbers.

4) Estimate the GMI as the empirical mean

$$R_a \approx \max_{s \geq 0} \frac{1}{S} \sum_{i=1}^{S} \log_2 \left( \frac{q_{\tilde{Y}|U}\left(\tilde{y}_k^{(i)}|u_k^{(i)}\right)^s}{\sum_{a \in \mathcal{U}} q_{\tilde{Y}|U}\left(\tilde{y}_k^{(i)}|a\right)^s \frac{1}{|\mathcal{U}|}} \right). \quad (19)$$

## IV. COMPLEXITY ANALYSIS

This section studies the complexity of MAGIQ over flat fading channels and compares it with SQUID [10, Sec. IV] and ADMM [11]. We start with a worst case analysis of Algorithm 1. Suppose the *while* loop is evaluated the maximum number $N$ of times. The first execution of the *while* loop evaluates (8) for each coordinate $n$ and each symbol from $\mathcal{X}$. This means that the product $\boldsymbol{Hx}$ must be performed $|\mathcal{X}|$ times, resulting in $NK|\mathcal{X}|$ multiplications. The norm evaluation requires another $NK|\mathcal{X}|$ multiplications and $NK^2|\mathcal{X}|$ additions. Finally, evaluating the minimum requires $\log(N|\mathcal{X}|)$ comparisons. We remark that these terms do not change inside the *while* loop. The further loop passes are thus dominated by additions and the minimum operation, giving a total worst case of $\sum_{i=1}^{N-1} \log((N-i)|\mathcal{X}|)$ comparisons. The worst case number of multiplications is thus $2NK|\mathcal{X}|$ with a memory with $NK|\mathcal{X}|$ entries.

The worst case complexity seems large, but the average complexity of MAGIQ can be substantially reduced. For example, one can initialize the algorithm with a good starting point $\boldsymbol{x}_{init}$. The results presented below are based on initializing MAGIQ with the QLP solution of the MF, which adds only $NK$ multiplications. One can further update a coordinate only if the cost reduction is significant. The significance level can be translated into a threshold for the norm update.

For example, Fig. 2 shows the empirical cumulative distribution function (CDF) of the total number of iterations of MAGIQ with the aforementioned fine tuning for $I = 6$ iterations for 16-QAM, over the whole SNR range $-10$ dB to 14 dB. The worst case number of loop passes is therefore $N \cdot I = 128 \cdot 6 = 768$, but the average number of loop passes is an order of magnitude lower than the worst case. Fig. 3 shows the GMI for 16 and 64-QAM for MAGIQ with and without the thresholding operation. We show that 64-QAM exhibits the largest gap in performance compared to no thresholding. The reduction in computational complexity comes at a price. However, even at the aggressive levels we have set it in this example, there is only a 0.5 dB gap at a spectral efficiency of 5.4 bpcu, corresponding to a code rate of 0.9, which is a reasonable operating point for a coded system. However, in the range 3 bpcu to 4.5 bpcu (corresponding to code rates of 0.5 to 0.75), the gap is insignificant. For 16-QAM there is virtually no penalty across the entire SNR range. The value of the threshold is $10^{-4}$ for 16-QAM and $10^{-6}$ for 64-QAM.

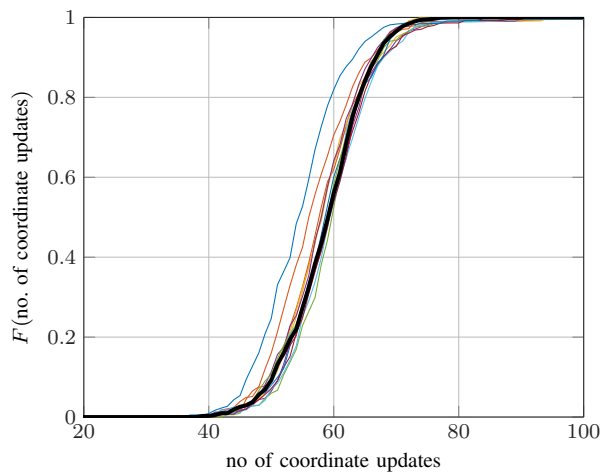

Fig. 2. Empirical CDF of the number of iterations with $N = 128$ and $I = 6$. The average CDF in thick black curve.

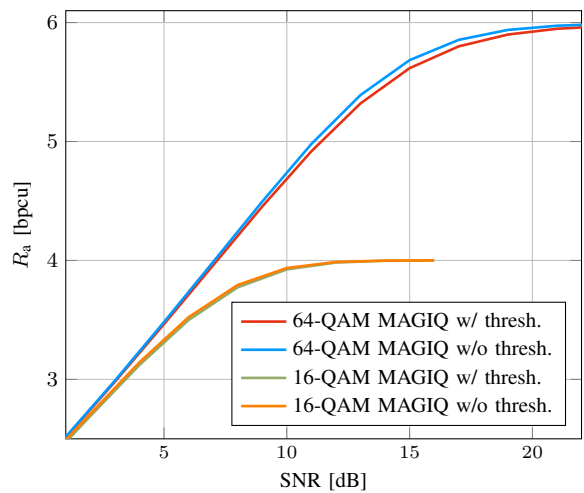

Fig. 3. GMI rates for $N = 128$ and $K = 16$.

The ADMM algorithm, in its least complex implementation (denoted as IDE2 in [11]), has a complexity of $4NK + 3N$ multiplications for the first iteration, and another $2NK + N$ multiplications for each new iteration. The reason is that initially computed quantities can be cached and then used as memory calls in later iterations. For SQUID there are $NK^2 + K^3 + 3NK$ multiplications in the pre-processing phase.

TABLE I
COMPUTATIONAL COMPLEXITY IN MULTIPLICATIONS AND COMPARISONS

| Algorithm | Total No. of multiplications | Total No. of comparisons |
|---|---|---|
| SQUID | $I \cdot (NK^2 + K^3 + 3NK) + I \cdot J \cdot (2NK + N)$ | $I \cdot J \cdot (N+1)\log(N)$ |
| SQUID Num. example Fig.5 | $1.01 \times 10^6$ | $1.8 \times 10^5$ |
| ADMM | $I \cdot (4NK + 3N) + I \cdot J \cdot (2NK + N)$ | $I \cdot J \cdot \log(N)$ |
| ADMM Num. example Fig.5 | $5.08 \times 10^5$ | 700 |
| MAGIQ (worst case) | $I \cdot (2NK|\mathcal{X}|) + NK + J \cdot 0$ | $I \cdot J \cdot \sum_{n=1}^{N-1} \log((N-n)|\mathcal{X}|)$ |
| MAGIQ Num. example Fig.5 | $6.34 \times 10^4$ | $1.07 \times 10^4$ |
| MAGIQ (average, $I=3$, $J_{average}=62$) Fig.2,3 | $6.34 \times 10^4$ | 568 |

After pre-processing, there are $2NK + N$ multiplications per iteration. If the precoding factor $\alpha$ is updated for ADMM and SQUID, we consider $I$ outer iterations that correspond to the number of times $\alpha$ is updated, and $J$ inner iterations to compute $x$ for a fixed value of $\alpha$.

Unlike MAGIQ that has a variable number of loop passes that depends on stopping criteria, ADMM and SQUID have fixed complexity once the number of iterations is fixed.

The per-sample complexity of MAGIQ for frequency selective channels is the same as that of MAGIQ for flat fading channels, but the metric is more involved due to convolutions. The metric computation for each of the $NK|\mathcal{X}|$ terms requires $L$ multiplications for the convolution. This results in a total of $NKL|\mathcal{X}|$ multiplications to retrieve the channel outputs. For the norm, again $NK|\mathcal{X}|L$ multiplications are required. This means that the per-sample complexity of MAGIQ increases by a factor equal to the channel length for the first iteration. In subsequent iterations the already computed convolutions can be reused by caching.

## V. NUMERICAL RESULTS

We first study the frequency-flat case. We compare the performance of different precoding schemes by means of their GMI for massive MIMO with $K = 16$ UEs and $N = 128$ antennas at the base station. If not stated otherwise all results are reported for Scenario 1 presented in Sec. III-E.

As a reference, we show the ZF solution with infinite precision ADCs, i.e., Q(·) in Sec. II-C is the identity function. We also include the performance of the MF and ZF QLP schemes, and the performance of the SQUID and ADMM algorithms. For the simulations, MAGIQ uses $I = 3$, SQUID uses $I = 50$ iterations, and ADMM uses $I \cdot J = 10 \cdot 10 = 100$ iterations. Going beyond these numbers of iterations did not result in further improvements. We do not claim that these numbers are optimal, but we tuned them for good performance for higher order modulations.

We apply the same $\alpha$ optimization as Scenario 1 to SQUID with $J = 4$ outer iterations. We call this precoding algorithm SQUID-$\alpha$. For MAGIQ and ADMM, the quantization resolution is one bit per real dimension, i.e., we set $b = 2$ in (2). The precoding solution for SQUID uses phase modulation only, i.e., the zero symbol is not included in the transmitter alphabet. Also, MAGIQ is initialized with the solution of the quantized MF, in order to speed up convergence. The results are obtained from 2000 channel realizations for flat fading, and 120 channel realizations for frequency selective fading.

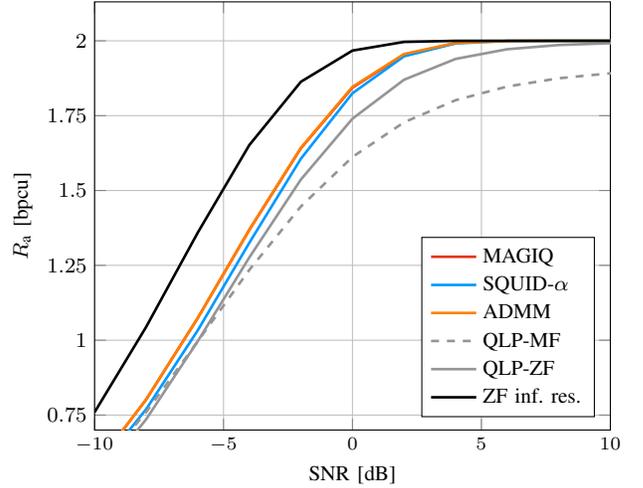

Fig. 4. GMI rates for QPSK, $N = 128$, and $K = 16$.

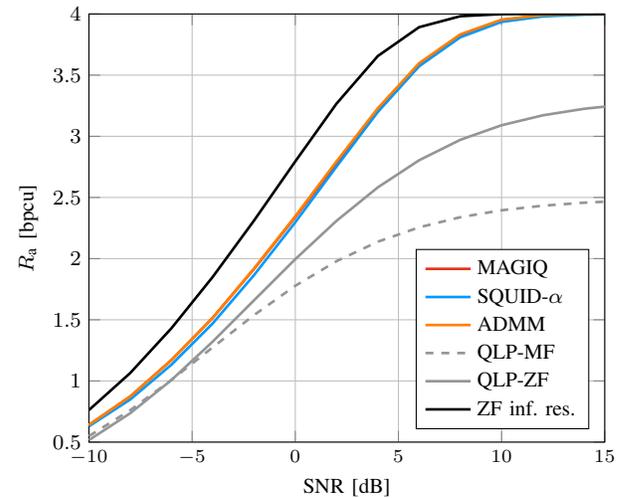

Fig. 5. GMI rates for 16-QAM, $N = 128$, and $K = 16$.

We display the achievable rates for QPSK, 16-QAM and 64-QAM in Figs. 4, 5 and 6, respectively. For QPSK modulation, SQUID-$\alpha$, ADMM, and MAGIQ show similar performance over the entire SNR range. QLP performs poorly at high SNR, and the MF solution even saturates below the maximum rate of 2 bpcu. QLP performs even worse for 16-QAM and 64-QAM. Both MAGIQ and ADMM show similar performance for higher order modulation formats. For 16-QAM, SQUID-$\alpha$ with adaptive $\alpha$ is competitive with MAGIQ and ADMM, but it does not reach the maximum rate at high SNR for 64-QAM.

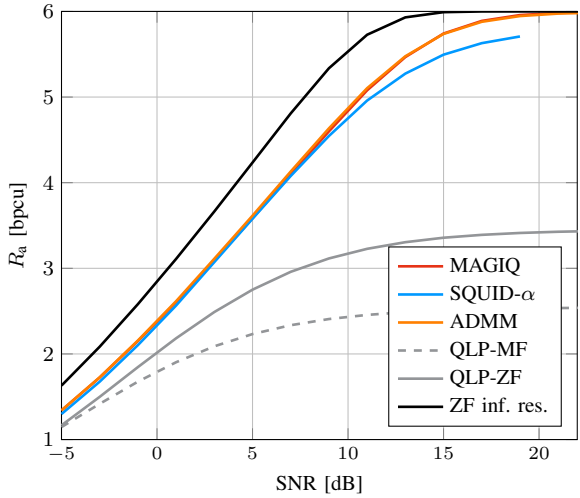

Fig. 6. GMI rates for 64-QAM, $N = 128$, and $K = 16$.

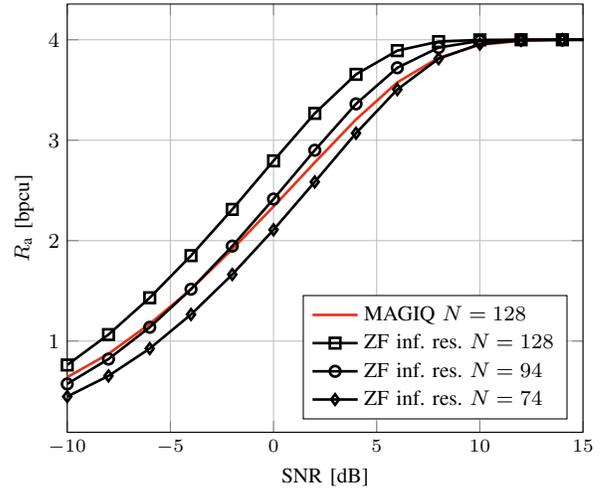

Fig. 8. GMI rates with infinite resolution ZF and different numbers of antennas for $K = 16$ and 16-QAM.

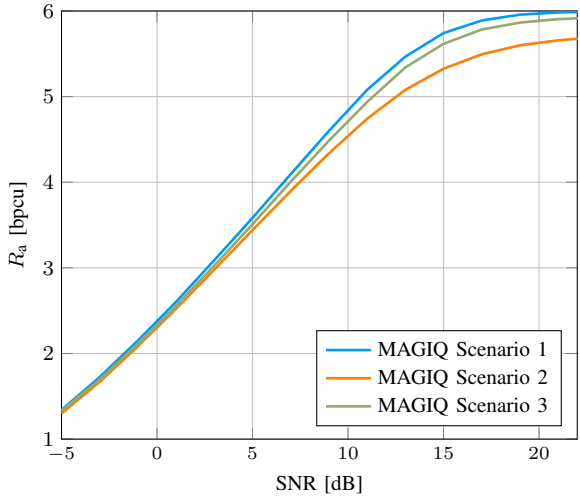

Fig. 7. GMI rates for 64-QAM, $N = 128$, and $K = 16$.

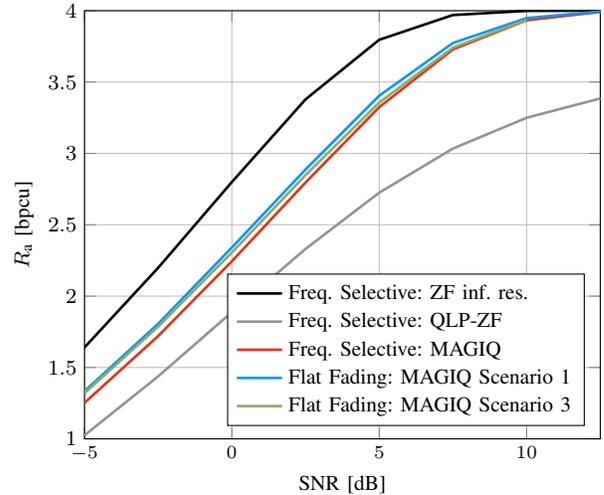

Fig. 9. GMI rates for OFDM, 16-QAM, $N = 128$, $K = 16$, $T = 286$, $L = 15$.

Figure 7 shows the achievable rates with 64-QAM in a block fading scenario with the transceivers designed according to the guidelines presented in III-E. Observe that both Scenarios 2 and 3 seem to be interference limited at high SNR, but the Scenario 3 operates close to Scenario 1. This demonstrates that knowing the precoding factors at the receiver may not be necessary.

Figure 8 compares the performance of MAGIQ with $K = 16$ users and 16-QAM to ZF with infinite precision for different numbers $N$ of antennas. Observe that the MAGIQ curve lies between the ZF performance for $N = 74$ and $N = 94$ antennas. A modest 1.5 to 1.7-fold increase in the number of antennas compensates for quantizing with $b = 2$.

Frequency selective fading is considered in Fig. 9 for a channel with $L = 15$ taps, $K = 16$ users, and OFDM symbols with $T = 286$.

We evaluate the GMI for 16-QAM with OFDM. The GMI reported is the average of the subcarrier rates. The frequency selective MAGIQ algorithm performs 4 iterations per OFDM symbol and computes a single precoding factor over the OFDM symbol. The algorithm was initialized with a time domain quantized solution of the frequency domain MF. The results show that the gap to the flat fading Scenario 1 solution is small.

## VI. CONCLUSION

We introduced MAGIQ for quantized precoding in a massive MIMO downlink channel. MAGIQ applies to both frequency-flat and frequency selective channels. Numerical performance comparisons using lower bounds on information rates suggests that MAGIQ outperforms QLP and SQUID, and it achieves similar performance as ADMM. We studied different update schedules of the precoding factor $\alpha$. For frequency-selective channels and OFDM, MAGIQ loses only about 0.25 dB as compared to the frequency-flat case for 16-QAM at 3.0 bpcu.

Future work will consider the combination of MAGIQ with different scheduling techniques and the influence of imperfect channel knowledge at the receiver. Another research direction may involve improved estimation of the decoding parameters, i.e., the parameters for the auxiliary channel and the precoding factor $\alpha$.